# A sealed ceramic GEM-based neutron detector

Jianjin Zhou [a,b,c], Jianrong Zhou [b,c,**], Xiaojuan Zhou [b,c], Lin Zhu [b,c], Yangdong Wei [d], Hong Xu [b,c], Huiyin Wu [e], Kang Wei [a], Jianqing Yang [b,c], Guian Yang [b,c], Yuguang Xie [c], Yi Zhang [a], Xiaohu Wang [d], Baowei Ding [a,*], Bitao Hu [a], Zhijia Sun [b,c,***], Yuanbo Chen [b,c]

a   School of Nuclear Science and Technology, Lanzhou University, Lanzhou, 730000, China
b   Spallation Neutron Source Science Center, Dongguan 523803, Guangdong, China
c   State Key Laboratory of Particle Detection and Electronics, Institute of High Energy Physics, Chinese Academy of Sciences, Beijing 100049, China
d   School of National Defense Science and Technology, Southwest University of Science and Technology, Mianyang 621000, China
e   Chongqing Jian'an Instrument Co., Ltd., Chongqing 400060, China

**Abstract**

The GEM-based neutron detector has flourished in the past decade. However almost all the GEM-based neutron detectors work in the flow-gas mode, and the long-term performances of the detectors may be unstable due to the dynamic changes of atmospheric pressure and ambient temperature. In this paper, a sealed ceramic GEM-based neutron detector was developed at China Spallation Neutron Source (CSNS) and its sensitive area was 100 mm × 100 mm. The characterizations of the detector were presented and discussed, which included the plateau curve, the neutron beam profile, the neutron wavelength spectrum, the spatial resolution (FWHM: 2.77±0.01(stat) mm), the two-dimensional (2D) imaging ability, the neutron detection efficiency and the counting rate instability (Relative Standard Deviation (RSD): 0.7%). The results show that the detector has good performances in sealed mode, and it can be used for the measurement of the direct neutron beam at CSNS.

*Corresponding author: Baowei Ding, dingbw@lzu.edu.cn, Lanzhou, China
**Corresponding author: Jianrong Zhou, zhoujr@ihep.ac.cn, Dongguan, China
***Corresponding author: Zhijia Sun, sunzj@ihep.ac.cn, Beijing, China

## 1. Introduction

CSNS [1] is a high-flux pulsed spallation neutron source which is similar to SNS in the USA [2], J-PARC in Japan [3], ISIS in the UK [4] and ESS in Europe [5] and it has been in public operation at 100kW power and 25Hz. At present, there are three neutron scattering instruments which have been built and seventeen which are being planned at CSNS. Efficient neutron detectors are an important component for every neutron scattering instrument. The dominant detectors are currently $^3$He-based gas detectors, such as Linear Position-Sensitive Detectors (LPSD) and Multi-Wire Proportional Chambers (MWPC). The counting rate of these detectors is a few tens kHz, which is limited by the space charge accumulation near the wire [6]. In addition, the price of $^3$He has been rising because of the $^3$He crisis in 2008 [7]. It is necessary to develop alternative neutron conversion materials and detector structures.

Boron-10 has a larger absorption cross section at energies of thermal (~$4\times10^3$ b) and epithermal (~$10^2$ b) neutrons than that of the other alternative neutron conversion materials such as Li. The boron converter layer made by evaporation or sputtering techniques is easily obtained, stable in physical and chemical properties, economic and environmental. The Gas Electron Multiplier (GEM) [8] detector, one of the micro-pattern gas detectors, offers excellent rate capability, good spatial resolution and time properties, radiation hardness and the possibility for large areas fabrication, and it is widely applied to high-energy physics. In the last decades, there was a booming development in GEM-based neutron detector by using boron conversion material [9]-[16]. The CASCADE neutron detector was firstly developed by Dr. Klein and Dr. Schmidt of Heidelberg University in Germany [9]. In addition to the standard GEM with 60 μm thickness, the GEM also derived many types, such as FR4 GEM with 400-800 μm thickness in Israel [17], ceramic GEM with 200 μm thickness in China [18]，Liquid Crystal Polymer (LCP) GEM [19] and Low Temperature Co-fired Ceramic (LTCC) GEM [20] with 100 μm thickness in Japan. At present, almost all the GEM-based neutron detectors work in flow-gas mode. Changes in atmospheric pressure and ambient temperature display daily and seasonal variations, the performances of detector may be unstable for long term. Meanwhile, the detector is not portable and not economical because of the gas system and gas consumption. In this paper, a sealed ceramic GEM-based neutron detector with a sensitive area of 100 mm × 100 mm was developed and the neutron irradiation test of this detector was carried out at CSNS.

**2. Detector setup**

The sealed ceramic GEM-based neutron detector had the similar structure with the flow-gas GEM-based detector. A schematic of detector was shown in Fig. 1. The detector was composed of boron-coated cathode, ceramic GEM and 2D signal readout board. The ions produced by the reaction of $^{10}$B(n, $\alpha$)$^7$Li generate primary electrons in drift region, the electrons are avalanched in GEM hole due to strong electric field, and the signals are induced in 2D readout board as a result of the movement of electrons in induction region. To ensure the long-term stable performances of the detector, on one hand, the gas volume of the detector must be isolated from the outside without gas exchange; on the other hand, the outgassing rate of the materials in the gas volume must be extremely low. The detailed specifications of some key components are elaborated.

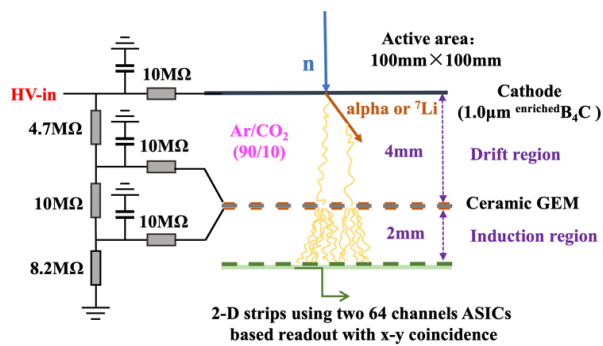

Fig. 1 Schematic of the detector.

The detector was a sandwich-like circular structure in the middle of which there was a stainless-steel flange used for welding high voltage feedthroughs, signal feedthroughs and gas tube. One side of the flange was the gas volume, and the other side was the electronics chamber. The gas volume was shown in Fig. 2(a), The high gas purity was maintained with the usage of ceramic materials for GEM, frames and screws which was a low-outgassing material. The cathode was a 300 μm thick aluminum (Al) plate which was coated with a 1.0 μm thick $B_4C$ ($^{10}B$ abundance: 96%). The ceramic GEM with 100 mm × 100 mm sensitive area was a double-copper coated ceramic foil of 200 μm thickness, the 200 μm diameter holes with a pitch of 600 μm were drilled and the 80 μm hole rims were etched by the printed circuit board technology. The ceramic GEM had higher transmittance and less self-scattering than FR4 GEM especially for cold neutrons, its performances were studied in Ref [18]. And the ceramic GEM had advantages including low outgassing, heatable and stable performances. The material of flexible readout board was Kapton and the thickness was about 0.6 mm. To obtain the neutron position, there were 128 cross-strip channels with a strip pitch of 1.56 mm on readout board, where 64 channels were assigned to x direction and 64 to y direction. The gas tube which was controlled by a needle valve was used for a vacuum environment and a pure gas filling. All components of the gas volume were sealed by an Al gasket, which ensured that the gas volume was isolated from the air. The electronics chamber was shown in Fig. 2(b). By the signal transfer board and the signal feedthroughs, the signals with 128 channels were connected to two CPix ASIC chips (64 channels) [21] developed by Heidelberg ASIC lab. There was a very fast readout electronics system, based on analogue front-end readout chips in combination with modern Field-Programmable-Gate-Array (FPGA) technology. It was realized through the coincidence of correlation of signals in the time and space domain, and the detailed algorithm of neutron event reconstruction was described in Ref [16]. The high voltage board provided a suitable working voltage for the cathode and the ceramic GEM through a resistive chain.

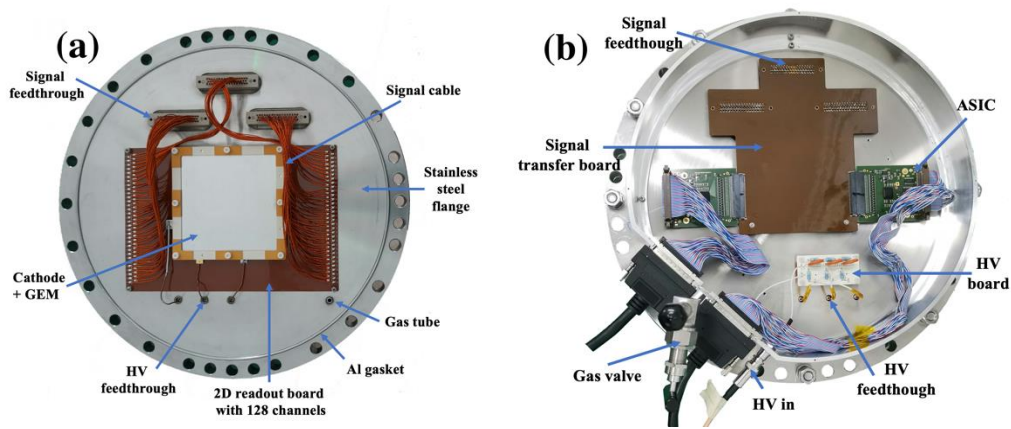

Fig. 2 (a) Photograph of gas volume. (b) Photograph of electronics chamber.

A special system was used for the detector outgassing: the detector was heated to 80°C to remove residual gas in the materials with heating tapes wrapped in Al foils for the heat uniformly. Under these conditions, the gas volume was vacuumed by the mechanical and molecular pumps for 5 days. During this period, the mixture gas of Ar and $CO_2$ (ratio 9:1, purity > 99.999%) was used to flush the gas volume for serval times. There were not hydrogen or fluoride components which were prone to ageing [22] in the mixture gas. Finally, the mixture gas was flushed into the gas volume until the pressure reached 1.0 atm.

**3. Experimental results**

The performances of the sealed ceramic GEM-based neutron detector, such as the plateau curve, the neutron beam profile, the neutron wavelength spectrum, the spatial resolution, the 2D imaging ability, the neutron detection efficiency and the counting rate stability, were studied using Time Of Flight (TOF) approach at BL20 of CSNS. The moderator was decoupled poisoned liquid hydrogen and the diameter of neutron beam exit was 20 mm in the BL20. A commercial neutron beam monitor, ORDELA 4562N (monitor), was placed at beam exit and measured simultaneously with the detector to eliminate neutron beam fluctuation in the experiments.

3.1 Plateau curve

The plateau curve was the counting rate curve with different working high voltage (HV). The counting rates were measured at HV from -1100V to -1800V to find the appropriate HV of the detector. Meanwhile, the counting rate of the monitor was also measured. The result was shown in Fig. 3, the normalized count was defined as the ratio of the counting rate of the detector to that of the monitor. The plateau slope of detector was 0.5%/100V with the plateau ranging from -1550V to -1800V. The fluctuation within the plateau range was caused by the low counts of the monitor. The detector HV was set at -1600V to improve the stability of the detector in the subsequent experiments.

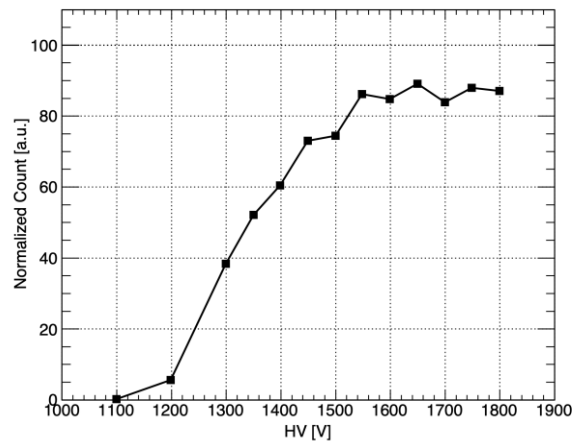

Fig. 3 Plateau curve of the detector. The plateau ranges from -1550V to -1800V and the plateau slope is 0.5%/100V. Normalized count is defined as the ratio of the counting rate of the detector to that of the monitor.

3.2 Neutron beam profile and neutron wavelength spectrum

The neutron beam profile and the neutron wavelength spectrum were measured using the TOF method. Limited by the size of the detector, it could not be placed close to the beam exit and placed 1.4 m away from the beam exit to measure the direct beam. Due to the diffusion of neutron beam, the measured diameter of beam profile was about 36 mm which was larger than that of the beam exit (shown in Fig. 4(a)). To evaluate the neutron wavelength spectrum, the wavelength spectrum that the monitor measured was used as a comparison. Both of the spectrums from 0.5 Å to 10 Å were calibrated with their detection efficiency calculated by Geant4 [23] and then normalized to the maximum counts (shown in Fig. 4(b)). The spectrum shapes measured by the monitor and the detector were consistent with each other, and the result validated the accuracy of the detector measurement. For the neutrons with wavelengths from 2 Å to 3 Å, the counts of detector were slightly lower than that of monitor. The Al incident window thickness of the monitor was 1 mm and that of the detector was 6 mm, and there are four Bragg edges from 2 Å to 3 Å for the Al element, leading to the stronger Bragg diffraction intensity and lower counts of the detector.

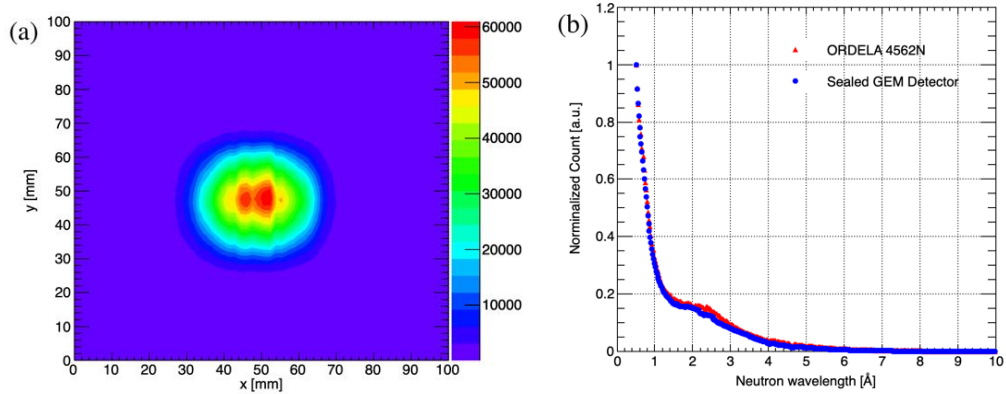

Fig. 4 (a) Imaging of neutron beam profile measured away from the beam exit 1.4 m at BL20 of CSNS. (b) Neutron wavelength spectrum of the detector compared with that of the monitor, both spectra were calibrated with detection efficiency and normalized to the maximum counts.

3.3 Spatial resolution and 2D imaging

Spatial resolution was an important parameter for 2D position-sensitive neutron detector. The spatial resolution could be measured by the Line Spread Function (LSF) method with a sufficiently narrow slit, and it was defined as the Full-Width-Half-Maximum (FWHM) of the peak in the projection on X-axis or Y-axis. A 2 mm-thick Cadmium (Cd) mask, with three slits of 0.5 mm wide and 20 mm pitch, was employed. The Cd mask was mounted in the font of the incident window of detector, and the spatial resolution of x and y directions were measured by rotating the Cd mask 90 degrees. In order to obtain the similar counts for the three slits, the detector was positioned at the back end of the beam line, about 4.96 m away from the beam exit. The detector responses were shown in Fig. 5, and the FWHM was obtained by Gaussian fitting. The distances between all peaks were about 20mm, which was consistent with that of the Cd mask. The spatial resolution in x direction was below 2.77±0.01(stat) mm (FWHM) and that in y direction was below 2.55±0.01(stat) mm (FWHM). The detector had good spatial resolution, and the spatial resolution could be further improved by reducing the pitch of strips in 2D readout board.

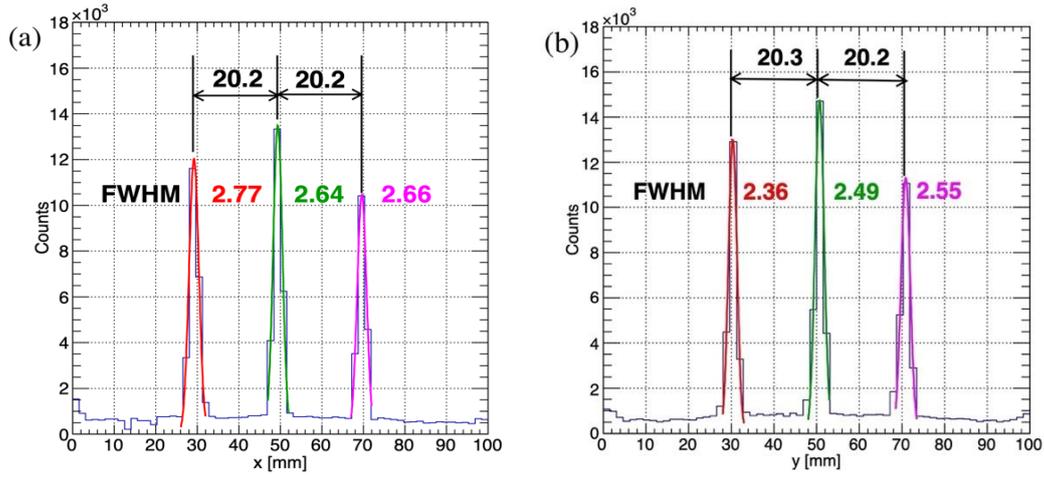

Fig. 5 Measurement of spatial resolution in X (a) and Y (b) directions. The Cd mask with three slits was employed, the slit width was 0.5mm and the pitch was 20mm.

2D imaging was measured by a 2 mm-thick Cd mask with a "CSNS" pattern (shown in Fig. 6(a)), and the imaging result was shown in Fig. 6(b). The four letters could be clearly distinguished which were consistent with the pattern and the background signal was low. In the imaging, the counts were high in the middle and low on the left and right sides, which was caused by the uneven neutron flux in the beam profile. The result indicated that the detector had a good capability of 2D imaging.

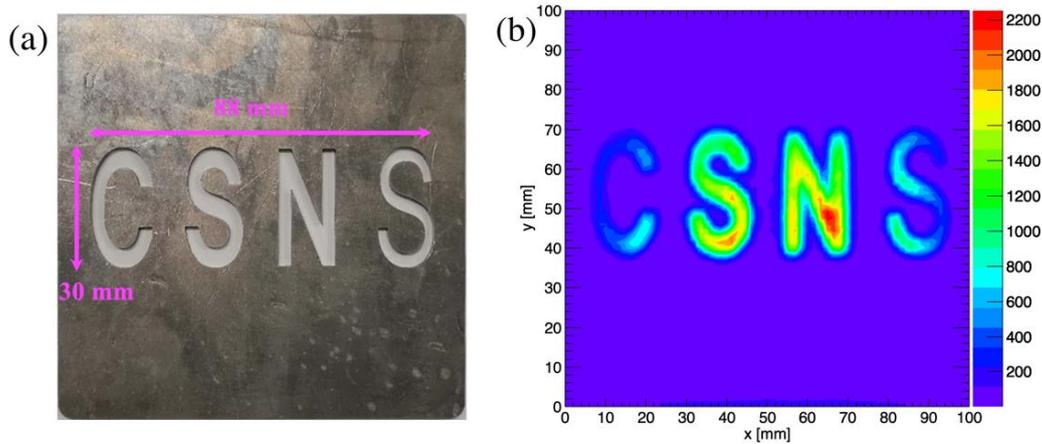

Fig. 6 (a) Cd mask with "CSNS" pattern, (b) the result of 2D imaging.

3.4 Neutron detection efficiency

The enriched $B_4C$ with 1.0 um thickness was employed as neutron converter. The neutron detection efficiency of the detector was evaluated by comparing with the standard $^3$He counting tube (20 atm, 1 inch in diameter). And it was the ratio of the normalized count of the detector to that of the $^3$He tube, the normalized count was defined as the ratio of the counts of detector or $^3$He tube to that of monitor. After being deflected by MICA monochromator, the neutron beam was collimated by a 5 mm thick boron aluminum alloy plate with a small hole (2 mm in diameter). The hole was aligned with the center of the $^3$He tube or the sensitive area of the detector. The neutron detection efficiency for 1.56 Å was measured to 2.3±0.1(stat)%. As a comparison, the relationship between the detection efficiency and the neutron wavelength was obtained by Geant4 (shown in Fig. 7). For the neutron with a wavelength of 1.56 Å, the simulation result of detection efficiency was 2.5%. The deviation between the results of experiment and simulation might be due to the thickness of $B_4C$ and incident window.

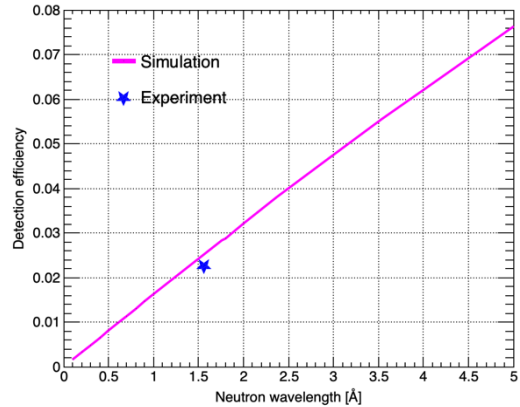

Fig. 7 Comparison of neutron detection efficiency between simulation and experiment. Neutron detection efficiency varies with neutron wavelength (calculated by Geant4, magenta curve). The experiment measurement (blue star).

3.5 Counting rate stability

The long-term stability is crucial for the sealed ceramic GEM-based neutron detector. In order to evaluate its counting rate stability, the detector was measured continuously for 17 hours and the data from the detector, the monitor and the proton targeting was collected every 20 minutes. The results were shown in Fig. 8. The counting rate of detector were normalized by the counting rate of monitor (blue triangle) and protons (red dot), and their instability fluctuations which were defined by RSD were 0.64% and 0.73%, respectively. Both results showed that the detector had a good long-term counting rate stability.

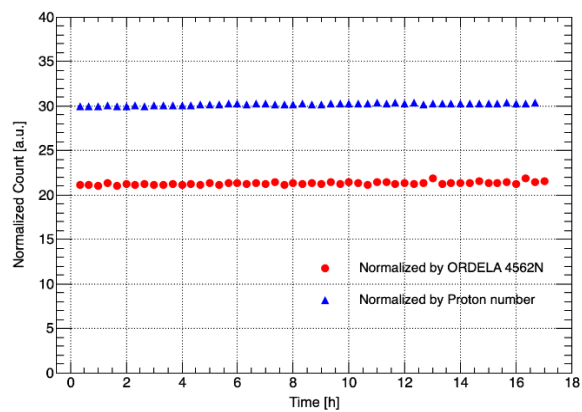

Fig. 8 Counting rate stability for 17 hours. The red dot represents the counts of the detector normalized by that of monitor, and the blue triangle represents the counts of the detector normalized by that of protons.

**4. Conclusion and outlook**

A sealed ceramic GEM-based neutron detector with a 100 mm×100 mm sensitive area was developed and measured at CSNS. The capability of 2D imaging and wavelength resolved was demonstrated. The plateau of detector ranged from -1550V to -1800V with 0.5%/100V slope, the spatial resolution was below 2.77±0.01 mm (FWHM), the detection efficiency for neutrons with a wavelength of 1.56 Å was 2.3±0.1%, the counts instability fluctuation was about 0.7% (RSD) in 17 hours of continuous measurement. The results show that the detector had good performances in sealed mode, and it could be used for neutron beam monitor, very small angle neutron scattering, neutron reflection and so on. In the future, a sealed cascade ceramic GEM-based neutron detector with high detection efficiency will be developed and used in the new instruments of CSNS.

**Acknowledgments**

This work was supported by the National Key R&D Program of China [Grant No. 2017YFA0403702], the National Natural Science Foundation of China [Grant No. 11635012, 11775243, U1832119], Youth Innovation Promotion Association CAS [Grant No. Y8291480K2] and Guangdong Basic and Applied Basic Research Foundation [2019A1515110217].